\documentclass[a4paper,12pt]{article}

\usepackage[T2A]{fontenc}
\usepackage[utf8]{inputenc}
\usepackage[russian,english]{babel}
\usepackage[sort,compress]{cite}
\usepackage{amsmath,amssymb}
\usepackage{bm}
\usepackage{geometry}
\usepackage{mathrsfs}
\usepackage{enumitem}

\allowdisplaybreaks[4]

\begin{document}

\begin{center}
  {
    \Large
    \bfseries
    Rarita--Schwinger field and multi-component wave equation\\[3ex]
  }

  A.E. Kaloshin$^{\dag}$ and V.P. Lomov$^{\dag\dag}$\\[2ex]
  {
    \small
    $^{\dag}$ Irkutsk State University\\
    $^{\dag\dag}$ The Institute for System Dynamics and Control Theory of SB RAS}
  \\[3ex]
  А.Е. Калошин$^{\dag}$, В.П. Ломов$^{\dag\dag}$\\[2ex]
  {\small
    $^{\dag}$ Иркутский государственный университет\\
    $^{\dag\dag}$ Институт динамики систем и теории управления СО РАН
  }
\end{center}

\begin{abstract}
  We suggest simple method to solve wave equation for Rarita--Schwinger field without additional
  constraints. This method based on use of off-shell projection operators allows to diagonalize
  spin-$1/2$ sector of the field.
\end{abstract}

\section*{Introduction}

It is well known that standard approach to description of free higher spin fields in quantum
field theory leads to constraints on field components (see,
e.g. \cite{Weinberg:1995})\footnote{Lagrange formulation of such theories needs to introduce
  auxiliary fields, see, e.g. \cite{Sorokin:2004ie,Vasiliev:2003e}}. But for interacting fields such
constraints can generate serious problems and contradictions. For Rarita--Schwinger field
\cite{Rarita:1941rs}, which is used for description of spin-3/2 particles, there are well-known
issues of such type, see \cite{Johnson:1961js,Velo:1969zw} and more general discussion in
\cite{Kobayshi:1987kt}.

One of the ways to avoid these problems with constrained fields is to choose the special form of
interaction to preserve constraints imposed on free field (if it's possible).

Another way is not to use the constraints at all and to eliminate the redundant components only
after calculating of the observables. It may be considered as some regularization.

Such an approach was used in \cite{Munczek:1967,Fukuyama:1973fy} for the free Rarita--Schwinger
field with all components: besides $s=3/2$ this field contains two extra $s=1/2$ components. It
was shown that it is possibly to regularly quantize free field
\cite{Munczek:1967,Fukuyama:1973fy} and that in this case the inclusion of electromagnetic
interaction does not generate the old contradictions \cite{Johnson:1961js,Velo:1969zw}. Follow
this line in \cite{Kaloshin:2004jh,Kaloshin:2007wc} we constructed the dressed propagator of
Rarita--Schwinger field with all components and found the renormalization procedure, which
guarantees the absence of redundant spin-1/2 poles. The essential point of our method is the use
of the off-shell projection operators.

Here we use this techniques to solve the wave equation for multi-component Rarita--Schwinger
field. It really gives simple and transparent method to solve multi-component equation as
compared with cumbersome calculations in \cite{Munczek:1967,Fukuyama:1973fy}. Moreover, in
spin-1/2 sector it allows to find explicit form of projectors onto mass states, which solves the
diagonalization problem.

\section{Wave equation for Rarita--Schwinger field}
\label{sec:wave_eq}

We use the action principle to get the wave equations for $\Psi_{\mu}$ field components. The
action for Rarita–Schwinger field is written in form
\begin{equation}
  \label{eq:rs_action}
  \mathscr{A} = \int\mathscr{L}\mathrm{d}^{4}x,\quad \mathscr{L}=\bar{\Psi}^{\mu}S_{\mu\nu}\Psi^{\nu},
\end{equation}
where $S_{\mu\nu}$ is a tensor-spinor operator. We write down it in form of $\Lambda$-basis
decomposition in momentum representation (see details in Appendix~\ref{sec:notations})
\begin{equation}
  \label{eq:L-decomp}
  S_{\mu\nu}=\sum_{i=1}^{10}\bar{S}_{i}\mathcal{P}_{\mu\nu}^{i},
\end{equation}
with arbitrary coefficients $\bar{S}_{i}$, where bar is used to denote coefficients of
$\Lambda$-basis.

The wave function $\Psi_{\mu}$ can be expressed as sum of orthogonal components
\begin{equation}
  \label{eeq:2}
  \Psi_{\mu}=n_{1\mu}\Psi_{1}+n_{2\mu}\Psi_{2}+\chi_{\mu},
\end{equation}
where $n_{i\mu}n^{\mu}_{j}=\delta_{ij}$, $n_{i\mu}\chi^{\mu}=0$, see Appendix.

The variation of the field may be written in the same form
$\delta\Psi_{\mu}=n_{1\mu}\delta\Psi_{1}+n_{2\mu}\delta\Psi_{2}+\delta\chi_{\mu}$
and $\delta\mathscr{L}$ is
\begin{equation*}
  \delta\mathscr{L}=\delta\bar{\Psi}_{1}n_{1\mu}(S^{\mu\nu}\Psi_{\nu})+
                    \delta\bar{\Psi}_{2}n_{2\mu}(S^{\mu\nu}\Psi_{\nu})+
                    \delta\bar{\chi}_{\mu}(S^{\mu\nu}\Psi_{\nu})+\mathrm{h.c.}.
\end{equation*}
Considering the $\delta\Psi_{i}$, $\delta\chi_{\mu}$ to be independent, after some algebra we
obtain system of equations for for $\Psi_{i}$, $i=1,2$:
\begin{equation}
  \label{eeq:5}
  \left\{
  \begin{array}{l}
    \Big(\bar{S}_{3}\Lambda^{-}+\bar{S}_{4}\Lambda^{+}\Big)\Psi_{1}+\Big(\bar{S}_{7}\Lambda^{-}+\bar{S}_{8}\Lambda^{+}\Big)\Psi_{2}=0,\\
    \Big(\bar{S}_{9}\Lambda^{+}+\bar{S}_{10}\Lambda^{-}\Big)\Psi_{1}+\Big(\bar{S}_{5}\Lambda^{+}+\bar{S}_{6}\Lambda^{-}\Big)\Psi_{2}=0,
  \end{array}\right.
\end{equation}
and single equation for $\chi_{\mu}$
\begin{equation}
  \label{eq:chi}
  \big(\bar{S}_{1}\Lambda^{+}+\bar{S}_{2}\Lambda^{-}\big)\chi_{\mu}=0.
\end{equation}
Here $\Lambda^{\pm}$ are off-shell projection operators $\Lambda^{\pm}=1/2(1\pm\hat{p}/W)$,
$W=\sqrt{p^{2}}$.

The equation \eqref{eq:chi} is usual Dirac equation for spin-$3/2$ particle, as it will be seen
below. The system \eqref{eeq:5} for $\Psi_{i}$ resembles the system of coupled Dirac
equations. It has non-trivial solutions only if $\Delta_{1}\Delta_{2}=0$, where
$\Delta_{1}=\bar{S}_{3}\bar{S}_{6}-\bar{S}_{7}\bar{S}_{10}$,
$\Delta_{2}=\bar{S}_{4}\bar{S}_{5}-\bar{S}_{8}\bar{S}_{9}$.

\section{Mass states in spin-$1/2$ sector}
\label{sec:projection-operators}

The system of equations \eqref{eeq:5} shows that $\Psi_{i}$ don't have the definite mass. In
order to diagonalize this system we use the projection representation of tensor-spinor operator
\begin{equation}
  \label{eq:3}
  S_{\mu\nu}=\sum_{i}\lambda_{i}\Gamma_{i\,\mu\nu},\quad
  (S^{-1})_{\mu\nu}=\sum_{i}\frac{1}{\lambda_{i}}\Gamma_{i\,\mu\nu},
\end{equation}
where $\Gamma_{i\mu\nu}$ are projection operators of $S_{\mu\nu}$ and $\lambda_{i}$
corresponding eigenvalues.

In our case it is convenient to formulate the eigenstate problem in terms of projectors.
So $\Gamma_{\mu\nu}$ should satisfy the following requirements
\begin{equation}
  \label{eq:13}
  S_{\mu}^{\;\alpha}\Gamma_{\alpha\nu}=\lambda\Gamma_{\mu\nu},\quad
  \Gamma_{i\,\mu}^{\;\;\;\alpha}\Gamma_{j\,\alpha\nu}=\delta_{ij}\Gamma_{i\,\mu\nu}.
\end{equation}

It is convenient to use $\Lambda$-basis decomposition for $\Gamma_{\mu\nu}$ to solve this
problem, see \eqref{eq:L-decomp}.

For spin-$3/2$ sector it is obvious that $\mathcal{P}_{1}$ and $\mathcal{P}_{2}$ are projection
operators and corresponding eigenvalues are $\bar{S}_{1}$ and $\bar{S}_{2}$.

For spin-$1/2$ terms we obtain the following equations on eigenvalues (here $E$ is a unit
matrix)
\begin{equation}
  \label{eq:14}
  \begin{split}
    \det(S_{1}-\lambda_{i}E)&=0,\; i=1,2;\quad S_{1}=
    \begin{pmatrix}
      \bar{S}_{3}  & \bar{S}_{7} \\
      \bar{S}_{10} & \bar{S}_{6}
    \end{pmatrix},\\[2ex]
    \det(S_{2}-\lambda_{j}E)&=0,\; j=3,4;\quad S_{2}=
    \begin{pmatrix}
      \bar{S}_{4} & \bar{S}_{8} \\
      \bar{S}_{9} & \bar{S}_{5}
    \end{pmatrix}.
  \end{split}
\end{equation}

The eigenvalues are $W$-dependent and are exchanged under the transformation $W\to-W$, namely
$1,2\leftrightarrow3,4$, that follows from the property of $\Lambda$-basis coefficients.

After some algebra we obtain four projection operators for spin-$1/2$ terms (tensor indices are
omitted)
\begin{align}
  \Gamma_{1} &= \dfrac{1}{\lambda_{2}-\lambda_{1}}\Big((\bar{S}_{6}-\lambda_{1})\mathcal{P}_{3}-\bar{S}_{10}\mathcal{P}_{10}\Big)+
    \dfrac{1}{\lambda_{2}-\lambda_{1}}\Big((\bar{S}_{3}-\lambda_{1})\mathcal{P}_{6}-\bar{S}_{7}\mathcal{P}_{7}\Big),\\
  \Gamma_{3} &= \dfrac{1}{\lambda_{4}-\lambda_{3}}\Big((\bar{S}_{5}-\lambda_{3})\mathcal{P}_{4}-\bar{S}_{9}\mathcal{P}_{9}\Big)+
    \dfrac{1}{\lambda_{4}-\lambda_{3}}\Big((\bar{S}_{4}-\lambda_{3})\mathcal{P}_{5}-\bar{S}_{8}\mathcal{P}_{8}\Big),\\
  \Gamma_{2} &= \dfrac{-1}{\lambda_{2}-\lambda_{1}}\Big((\bar{S}_{6}-\lambda_{2})\mathcal{P}_{3}-\bar{S}_{10}\mathcal{P}_{10}\Big)+
    \dfrac{-1}{\lambda_{2}-\lambda_{1}}\Big((\bar{S}_{3}-\lambda_{2})\mathcal{P}_{6}-\bar{S}_{7}\mathcal{P}_{7}\Big),\\
  \Gamma_{4} &= \dfrac{-1}{\lambda_{4}-\lambda_{3}}\Big((\bar{S}_{5}-\lambda_{4})\mathcal{P}_{4}-\bar{S}_{9}\mathcal{P}_{9}\Big)+
    \dfrac{-1}{\lambda_{4}-\lambda_{3}}\Big((\bar{S}_{4}-\lambda_{4})\mathcal{P}_{5}-\bar{S}_{8}\mathcal{P}_{8}\Big),
\end{align}
corresponding to eigenvalues \eqref{eq:14}.

The found projection operators allow us to obtain corresponding eigenvectors $l_{i\mu}$ which
have the following form
\begin{flalign}
  l_{1\mu} &= \sqrt{\frac{\bar{S}_{7}\bar{S}_{10}}{\bar{S}_{7}\bar{S}_{10}+|\bar{S}_{3}-\lambda_{1}|^{2}}}
    \bigg(n_{1\mu}-\frac{\bar{S}_{3}-\lambda_{1}}{\bar{S}_{7}}n_{2\mu}\bigg)\Lambda^{-},\\
  l_{2\mu} &= \sqrt{\frac{\bar{S}_{7}\bar{S}_{10}}{\bar{S}_{7}\bar{S}_{10}+|\bar{S}_{3}-\lambda_{2}|^{2}}}
    \bigg(n_{1\mu}-\frac{\bar{S}_{3}-\lambda_{2}}{\bar{S}_{7}}n_{2\mu}\bigg)\Lambda^{-},\\
  l_{3\mu} &= \sqrt{\frac{\bar{S}_{8}\bar{S}_{9}}{\bar{S}_{8}\bar{S}_{9}+|\bar{S}_{4}-\lambda_{3}|^{2}}}
    \bigg(n_{1\mu}-\frac{\bar{S}_{4}-\lambda_{3}}{\bar{S}_{8}}n_{2\mu}\bigg)\Lambda^{+},\\
  l_{4\mu} &= \sqrt{\frac{\bar{S}_{8}\bar{S}_{9}}{\bar{S}_{8}\bar{S}_{9}+|\bar{S}_{4}-\lambda_{4}|^{2}}}
    \bigg(n_{1\mu}-\frac{\bar{S}_{4}-\lambda_{3}}{\bar{S}_{8}}n_{2\mu}\bigg)\Lambda^{+}.
\end{flalign}
As a result the projection operators can be written as
\begin{equation}\label{eq:18}
  \Gamma_{i\mu\nu}=l_{i\mu}\bar{l}_{i\nu},
\end{equation}
where $\bar{l}_{\mu}=\gamma^{0}l_{\mu}^{\dag}\gamma^{0}$. The main property of $l_{i\mu}$ is
\begin{equation}
  \label{eq:10}
  \bar{l}_{i\mu}l^{\mu}_{i} = \Lambda^{-},\; i=1,2,\quad
  \bar{l}_{j\mu}l^{\mu}_{j} = \Lambda^{+},\; j=3,4.
\end{equation}

As the result the projection representation \eqref{eq:3} for $(S^{-1})_{\mu\nu}$ now takes more
concrete form
\begin{equation}
  \label{eq:4}
  (S^{-1})_{\mu\nu}=\frac{1}{\bar{S}_{1}}\mathcal{P}_{1\mu\nu}+\frac{1}{\bar{S}_{2}}\mathcal{P}_{2\mu\nu}+\sum_{i=1}^{4}\frac{1}{\lambda_{i}}\Gamma_{i\,\mu\nu}.
\end{equation}

Now it is possible to write wave function $\Psi_{\mu}$ as a sum of orthogonal components with
definite mass
\begin{equation}
  \label{eq:15}
  \Psi_{\mu}(p)=(l_{1\mu}+l_{3\mu})\varphi_{1}(p)+(l_{2\mu}+l_{4\mu})\varphi_{2}(p)+\chi_{\mu}(p),
\end{equation}
where $\chi_{\mu}l_{i}^{\mu}=0$, $i=1,\dots,4$, so Lagrangian \eqref{eq:rs_action} is expanded
into
\begin{equation}
  \label{eq:lagr-ev-expanded}
  \mathscr{L}=\bar{\chi}_{\mu}\big(\bar{S}_{1}\Lambda^{+}+\bar{S}_{2}\Lambda^{-}\big)\chi^{\mu}+
  \bar{\varphi}_{1}\big(\lambda_{3}\Lambda^{+}+\lambda_{1}\Lambda^{-}\big)\varphi_{1}+
  \bar{\varphi}_{2}\big(\lambda_{4}\Lambda^{+}+\lambda_{2}\Lambda^{-}\big)\varphi_{2}.
\end{equation}

Repeating the steps in deriving the wave equation but using instead of \eqref{eeq:2} the
decomposition \eqref{eq:15} we get independent motion equations for $\varphi_{1}$ and $\varphi_{2}$:
\begin{equation}
  \label{eq:19}
  \begin{array}{c}
    (\Lambda^{+}\lambda_{3}+\Lambda^{-}\lambda_{1})\varphi_{1}=0,\\
    (\Lambda^{+}\lambda_{4}+\Lambda^{-}\lambda_{2})\varphi_{2}=0.
  \end{array}
\end{equation}

\section{Lagrangian of massive Rarita--Schwinger field}
\label{sec:expl-form-eigenv}

To concretize the above general formulae, let's consider the most general form of Lagrangian of
free Rarita--Schwinger field\footnote{The most general Lagrangian depends on four parameters
  \cite{Kaloshin:2005wb, Munczek:1967, Pilling:2004cu} but poles positions depend only on two
  parameters. So without loosing generality we use two-parameter Lagrangian. Details can be
  found in \cite{Kaloshin:2005wb}.}
\begin{equation}
  \label{eq:12}
  \begin{aligned}
    \mathscr{L} &= \bar{\Psi}_{\mu}S^{\mu\nu}\Psi_{\nu},\\
    S^{\mu\nu}   &= g^{\mu\nu}(\hat{p}-M) - p^{\mu}\gamma^{\nu} -
                   p^{\nu}\gamma^{\mu}+\gamma^{\mu}\gamma^{\nu}M(1+r) + \gamma^{\mu}\hat{p}\gamma^{\nu}\big(\frac{\delta}{2}+1\big),
  \end{aligned}
\end{equation}
which besides mass $M$ has two real parameters $r$, $\delta$. Eigenvalues \eqref{eq:14} in
this case take the form
\begin{align}
  \label{eq:eigenv-l}
  \lambda_{1,2} &= M(1+2r)-W(1+\delta)\mp\Big[4M^{2}(1+r)^{2}-
                  2M(1+r)(1+2\delta)W+\big(1+\delta+\delta^{2}\big)W^{2}\Big]^{1/2},\\
  \label{eq:eigenv-e}
  \lambda_{3,4} &= M(1+2r)+W(1+\delta)\mp\Big[4M^{2}(1+r)^{2}+
                  2M(1+r)(1+2\delta)W+\big(1+\delta+\delta^{2}\big)W^{2}\Big]^{1/2}.
\end{align}
It is obvious that the values depend on $W$ and exchanged when $W\to-W$.

In general case eigenvalues $\lambda_{i}(W)$ have specific dependence on $W$, so equations
\eqref{eq:19} are not exactly Dirac equations. To see it, we can look again at system of
equations \eqref{eeq:5}. Indeed, we can rewrite the system for $\Psi_{i}$ as system of equations
resembling Dirac equation
\begin{equation}
  \label{eeq:11}
  \Big[\hat{p}\bm{K}-\bm{M}\Big]
  \begin{pmatrix}
    \Psi_{1} \\
    \Psi_{2}
  \end{pmatrix}
  =0
\end{equation}
with non-diagonal kinetic $\bm{K}$ and mass $\bm{M}$ matrices. The Lagrangian \eqref{eq:12}
leads to following matrices
\begin{align*}
  \bm{K}&=\frac{M}{W}
  \begin{pmatrix}
    0 & \sqrt{3}(1+r)\\
    \sqrt{3}(1+r) & 0
  \end{pmatrix}+
  \begin{pmatrix}
    (4+3\delta)/2 & 0\\
    0 & \delta/2
  \end{pmatrix},\\
  \bm{M}&=-M
  \begin{pmatrix}
    2+3r & 0 \\
    0 & r
  \end{pmatrix}-
  W
  \begin{pmatrix}
    0 & \sqrt{3}/2\delta\\
    \sqrt{3}/2\delta & 0
  \end{pmatrix}.
\end{align*}
Note that matrices $\bm{M}$ and $\bm{K}$ are $W$-dependent.

Zeroes of eigenvalues are poles for the propagator. The first two terms in \eqref{eq:4} have
poles in points $M$ and $-M$ correspondingly. Denote the poles in spin-$1/2$ sector as $m_{1}$,
$m_{2}$, i.e.
\begin{equation}
  \label{eq:5}
  \lambda_{1}(m_{1})=0,\quad \lambda_{2}(m_{2})=0.
\end{equation}
The above mentioned property of eigenvalues suggests that $-m_{1}$ and $-m_{2}$ are zeroes of
$\lambda_{3}$ and $\lambda_{4}$ correspondingly. From explicit form of eigenvalues
\eqref{eq:eigenv-l} one derive that
\begin{equation}
  \label{eq:6}
  m_{1,2}=M\frac{r-\delta\pm\sqrt{(\delta-r)^{2}+(3+4r)\delta}}{\delta}.
\end{equation}

Dirac case corresponds to particular choice of parameter $r=-1$, when eigenvectors are linear on
$W$:
\begin{align}
  \lambda_{1,2}&=-M-W(1+\delta\pm\sqrt{1+\delta+\delta^{2}}),\\
  \lambda_{3,4}&=-M+W(1+\delta\pm\sqrt{1+\delta+\delta^{2}}).
\end{align}
For this case equations \eqref{eq:19} have standard Dirac form and Lagrangian
\eqref{eq:lagr-ev-expanded} is written as
\begin{equation}
  \label{eq:lagr-ev-exp-explicit}
  \mathscr{L}=\bar{\chi}_{\mu}\big(\bar{S}_{1}\Lambda^{+}+\bar{S}_{2}\Lambda^{-}\big)+
    \bar{\varphi}_{1}\big((1+\delta+\sqrt{1+\delta+\delta^{2}})\hat{p}-M\big)\varphi_{1}+
    \bar{\varphi}_{2}\big((1+\delta-\sqrt{1+\delta+\delta^{2}})\hat{p}-M\big)\varphi_{2}.
\end{equation}

The diagonal form of Lagrangian \eqref{eq:lagr-ev-exp-explicit} allows to see sign of every
contributions to the energy. Indeed, comparing with Dirac case, we see that
field $\varphi_{1}$ gives always positive contribution to Hamiltonian, while contribution of
$\varphi_{2}$ depends on sign of $\delta$; if $\delta$ is negative then Hamiltonian has negative
contribution from $\varphi_{2}$. This conclusion differs from conclusion obtained in work
\cite{Munczek:1967}: according to it at least one contribution from spin $1/2$ components to
Hamiltonian is negative.

\section*{Conclusions}

We found that with use of $\Lambda$-basis \eqref{eq:L-decomp} and decomposition of field
\eqref{eeq:2}, it's easy to obtain the wave equations for different spin sectors of
Rarita–Schwinger field. For spin-3/2 it is in fact Dirac equation, as for spin-1/2 sector, we
have two coupled Dirac-like equations with non-diagonal kinetic and mass matrices. In general
case the mass matrix in \eqref{eeq:11} is energy dependent $M(W)$, only special choose of
parameters leads to constant matrix.

We used the most general form of wave operator \eqref{eq:L-decomp} because the above presented
Lagrangians \eqref{eq:12} don't exhaust all possibilities. For instance, there exist some
examples of Rarita–Schwinger Lagrangians with higher derivatives
\cite{Kruglov:2004iz,Rico:2007br}.

$\Lambda$-basis allows also to find projectors onto the mass states for spin-1/2
sector. Convenient trick here is the use of projection representation of operator in form of
\eqref{eq:3}, \eqref{eq:15}.

We can conclude that presented here methods allow to work effectively with multi-component field
without additional constraints and may be useful for other higher spin fields and more complicated
Lagrangians.

\section*{Acknowledgments}

This work was supported in part by the program «Development of Scientific Potential in Higher
Schools» (project 2.2.1.1/1483, 2.1.1/1539) and by the Russian Foundation for Basic Research
(project No. 09-02-00749).

\appendix

\section{Notations}
\label{sec:notations}

Throughout the work we use the following notation. The elements of $\Lambda$-basis are defined
as
\begin{equation} \label{eq:L-basis}
  \begin{aligned}
    \mathcal{P}_{1} &= \Lambda^{+}\mathcal{P}^{3/2}, & \mathcal{P}_{2} &= \Lambda^{-}\mathcal{P}^{3/2}, &
    \mathcal{P}_{3} &= \Lambda^{+}\mathcal{P}^{1/2}_{11}, & \mathcal{P}_{4} &= \Lambda^{-}\mathcal{P}^{1/2}_{11},\\
    \mathcal{P}_{5} &= \Lambda^{+}\mathcal{P}^{1/2}_{22}, & \mathcal{P}_{6} &= \Lambda^{-}\mathcal{P}^{1/2}_{22}, &
    \mathcal{P}_{7} &= \Lambda^{+}\mathcal{P}^{1/2}_{21}, & \mathcal{P}_{8} &= \Lambda^{-}\mathcal{P}^{1/2}_{21}, \\
    && \mathcal{P}_{9} &= \Lambda^{+}\mathcal{P}^{1/2}_{12}, & \mathcal{P}_{10} &= \Lambda^{-}\mathcal{P}^{1/2}_{12}, & &
  \end{aligned}
\end{equation}
where
\begin{gather} \label{eq:oper}
  \begin{split}
    (\mathcal{P}^{3/2})^{\mu\nu}&=g^{\mu\nu}-n_{1}^{\mu}n_{1}^{\nu}-n_{2}^{\mu}n_{2}^{\nu}\\
    (\mathcal{P}^{1/2}_{11})^{\mu\nu}&=n_{1}^{\mu}n_{1}^{\nu}, \qquad
    (\mathcal{P}^{1/2}_{22})^{\mu\nu}=n_{2}^{\mu}n_{2}^{\nu},  \\
    (\mathcal{P}^{1/2}_{21})^{\mu\nu}&=n_{1}^{\mu}n_{2}^{\nu}, \qquad
    (\mathcal{P}^{1/2}_{12})^{\mu\nu}=n_{2}^{\mu}n_{1}^{\nu}.
  \end{split}
\end{gather}
The «unit» vectors $n_{1\mu}$ and $n_{2\mu}$ are defined as following
\begin{equation}
  \label{eq:9}
  n_{1\mu}=\frac{1}{\sqrt{3}}\Big(g_{\mu\alpha}-\frac{p_{\mu}p_{\alpha}}{p^{2}}\Big)\gamma^{\alpha},\quad
  n_{2\mu}=\frac{p_{\mu}}{\sqrt{p^{2}}},\quad
  n_{i\mu}\cdot n^{\mu}_{j}=\delta_{ij},
\end{equation}
and
\begin{equation}
  \label{eq:11}
  \Lambda^{\pm}=\frac{1}{2}\Big(1\pm\frac{\hat{p}}{W}\Big),\quad W=\sqrt{p^{2}}.
\end{equation}
The decomposition over the $\Lambda$-basis of $S_{\mu\nu}$ is written as
\begin{equation}
  \label{eq:8}
  S_{\mu\nu}=\sum_{i=1}^{10}\bar{S}_{i}\mathcal{P}_{\mu\nu}^{i}.
\end{equation}



\end{document}